Modeling cosmic ray proton induced terrestrial neutron flux: A lookup table

Andrew C. Overholt[1,2], Adrian L. Melott[1], and Dimitra Atri[3,4]

[1] Department of Physics and Astronomy, University of Kansas, 1251 Wescoe Hall Dr. #1082, Lawrence, Kansas 66045
[2] Department of Science and Mathematics, MidAmerica Nazarene University, 2030 East College Way, Olathe, Kansas 66062
[3] Department of High Energy Physics, Tata Institute of Fundamental Research, Homi Bhabha Road, Mumbai 400 005, India
[4] Blue Marble Space Institute of Science, Seattle, WA 98145-1561

ABSTRACT

Neutrons contribute a significant radiation dose at commercial passenger airplane altitudes. With cosmic ray energies $> 1\ GeV$, these effects could in principle be propagated to ground level. Under current conditions, the cosmic ray spectrum incident on the Earth is dominated by particles with energies $< 1\ GeV$. Astrophysical shocks from events such as supernovae accelerate high energy cosmic rays (HECRs) well above this range. The Earth is likely episodically exposed to a greatly increased HECR flux from such events. Solar events of smaller energies are much more common and short lived, but still remain a topic of interest due to the ground level enhancements (GLEs) they produce. The air showers produced by cosmic rays (CRs) ionize the atmosphere and produce harmful secondary particles such as muons and neutrons. Although the secondary spectra from current day terrestrial cosmic ray flux are well known, this is not true for spectra produced by many astrophysical events. This work shows the results of Monte Carlo simulations quantifying the neutron flux due to CRs at various primary energies and altitudes. We provide here lookup tables (described later) that can be used to determine neutron fluxes from proton primaries with total energies $1\ MeV - 1\ PeV$. By convolution, one can compute the neutron flux for any arbitrary CR spectrum. This contrasts with all other similar work, which is spectrum dependent. Our results demonstrate the difficulty in deducing the nature of primaries from the spectrum of ground level neutron enhancements.

1. INTRODUCTION

Cosmogenic neutrons and their effects are a large area of research. This research includes atmospheric neutron measurements which have been taken since the 1950s at a variety of latitudes and altitudes [*Davis* 1950, *Hess et al.* 1959, *Gordon et al.* 2004, *Goldhagen et al*. 2004, *Moser et al.* 2005, among others]. These measurements are useful for determination of cosmic ray flux variability. As these measurements span only decades, they are inadequate for long term events such as nearby supernovae, and are primarily used to examine solar events. Shorter events such as solar proton events (SPEs) and the ground level enhancements (GLEs) of neutrons they produce are an area of ongoing research [*Duldig et al.*, 1994; *Gordon et al.*, 2004; *Gopalswamy et al.,* 2005; among others]. An SPE bombards the Earth with high energy protons; the upper limit on the total energy of Solar SPEs is not known [*Melott and Thomas*, 2012, *Thomas et al*. 2013]. The cosmic rays produced in these events are of sufficient energy for shower production within our atmosphere [*O'Brien et al.*, 1996].

The Earth is constantly bombarded by high energy particles referred to as cosmic rays. These particles are primarily protons with a small component of helium nuclei, heavier nuclei, and electrons. These primary particles interact with air molecules. Target nuclei often undergo spallation, dividing them into smaller nuclei and constituent nucleons. These energetic secondary particles produce additional interactions as they continue to propagate through our atmosphere. This cascade of particles contains a soft component (e.g. photons and charged nuclei), and a hard component (e.g. muons and neutrons). Most of the energy of the primary particle is expended on atmospheric ionization, which creates the majority of the soft component. Due to its larger interaction cross section, this component rarely reaches ground level. Therefore this work focuses on the hard component of cosmic ray showers.

Muons are primarily produced through the decay of pions which are created in high energy interactions with target nucleons. Due to their high penetration, muons are capable of reaching the ground, penetrating deep underground, or into water. This component of cosmic ray showers has been calculated in previous work [*Atri and Melott*, 2011a, 2011b]. Although muons have a small biological effectiveness, they contribute significantly to the background radiation dose due to their large abundance at ground level. This contribution will be greatly increased during a HECR event, impacting terrestrial biota [*Atri and Melott* 2011b, 2013].

Neutrons are formed through the process of cosmic ray spallation when high energy particles collide with atmospheric nuclei. Neutrons penetrate much further into our atmosphere than the electromagnetic component as they are not geomagnetically trapped. As they propagate, they collide with atmospheric nuclei, liberating additional particles as well as slowing to thermal energies (~0.025 *eV*). Once at thermal speeds, these neutrons exhibit only random motion due to their energy. These thermal neutrons are typically deposited in the stratosphere for lower energy primaries, with higher energy primaries producing thermal neutrons closer to ground level (but still at altitude) as shown in Figure 1.

Neutrons constitute a source of cosmic radiation due to their high biological effectiveness [*Reitz*, 1993]. They are a substantial risk factor at high latitudes in the stratosphere under current day conditions [*e. g. Reitz*, 1993; *O'Brien et al*., 1996; *Kojo et al*., 2005; *Hammer et al*., 2009; *Beck*, 2009]. Cancer rates and rates of spontaneous abortion are increased significantly among flight crews [*Pukkala et al.*, 1995; *Aspholm et al.,* 1999]. Variability in galactic cosmic ray flux has been linked to increased cancer mortality and breast cancer [*Juckett and Rosenberg*, 1997; *Juckett*, 2007, 2009].

Although solar events have been studied via neutron monitor measurements, no measurements exist for cosmic ray spectra produced by other astrophysical events. There is a non-trivial probability of events such as nearby supernovae [*Erlykin and Wolfendale*, 2001; *Fields et al.,* 2008] and gamma ray bursts (GRBs) [*Dermer and Holmes*, 2005; *Kusenko*, 2010] exposing the Earth to an enhanced flux of cosmic rays over ~ 100 *Myr* timescales [*Erlykin and Wolfendale* 2010; *Melott and Thomas*, 2011]. Motion of the sun perpendicular to the galactic plane has also been proposed to increase HECR flux due to increased exposure to the galactic shock [*Medvedev and Melott*, 2007]. These events produce substantially higher primary energy than SPEs, making them capable of devastating direct effects. Unlike other work which has focused on one case, our work is applicable to any arbitrary HECR event, if the cosmic ray spectrum for the event is known. These events also produce high-energy photons which have been modeled in detail [*Thomas et al.,* 2005; *Ejzak et al.,* 2007] and will not be discussed further as we focus on cosmic ray effects.

Monte Carlo simulations have been performed for typical current cosmic ray fluxes, and have shown to be reliable under these circumstances [*Goldhagen et al.*, 2004; *Grigoriev et al.*, 2010]. Such work has been used to study ground level enhancements (GLEs) of neutron flux [*Duldig et al.*, 1994; *Gordon et al.*, 2004; *Gopalswamy et al.,* 2005; among others]. These enhancements are indicative of solar proton events (SPEs), and have been used to study their properties through the use of ground-based and atmospheric neutron monitors. Although this work is extensive, no work has been done to provide neutron fluxes independent of primary spectrum or for energies in the *TeV* range and above. For this reason we have tabulated neutron fluxes for a wide range of primary energies. These results can be used to simulate the resulting neutron component in the shower from arbitrary cosmic ray spectra. Lookup tables represent proton primaries with kinetic energies ranging from $10^6$ to $10^{15}$ *eV*. This is different from representing it as a function of the total energy of the particle, which is standard in high energy physics. The total energy can be found by adding the rest mass energy of the proton (9.38272 x $10^8$ *eV*) to our numbers.

Protons comprise the large majority of cosmic rays and thus can be used as an approximation for all cosmic ray showers. We extend this by representing heavier nuclei by ensembles of protons with the given atomic weight. Under this approximation, results still slightly underestimate cosmic ray showers containing heavy nuclei. As other nuclei compose ~10% of cosmic ray primaries, this should be an adequate approximation despite the larger interaction cross section of heavy nuclei. We can estimate the size of the effect as follows: Only about 1% of cosmic rays are heavier than alpha particles; alphas comprise about 10%. We have verified with some tests that at high energies such nuclei can be approximated as the sum of the number of nucleons with a small error. For alphas, this error is greatest, about 50%, below 100 MeV and much smaller for higher energies. We can thus put an upper limit on the total systematic underestimate of neutron production of 5% from this, which is smaller than statistical error (about 20% below 100 *Mev*).

Protons below $10^6$ *eV* are not simulated in our work, as the neutron production cross section is insignificant below a few *MeV*. Our table therefore represents the largest portion of relevant cosmic rays, and is therefore suitable for application to a wide range of astrophysical sources.

2. COMPUTATIONAL MODELING

Simulations were run as a two-step process. CORSIKA (COsmic Ray SImulations for KAscade) [*Heck*, 2001] was used for high energy interactions ($> 10$ *GeV*), while MCNP [*Brown et al.*, 2002] and MCNPx [*Pelowitz*, 2005] were used for neutron thermalization and propagation (neutrons below 50 *MeV*) as well as low primary energy shower formation ($< 10$ *GeV*). We calculated particle fluxes from single proton primaries, creating tables which can be used to deduce results from many different spectra.

CORSIKA is a Monte Carlo code used extensively to study air showers generated by primaries up to 100 *EeV*. It is well suited to high energy interactions, as it is calibrated using KASCADE, a detector used to study hadronic interactions in the $10^{16}$ to $10^{18}$ *eV* energy range, as well as with a number of other experiments around the globe. CORSIKA 6.960 was used for all high energy simulations. The code was set up with EPOS as the high-energy hadronic interaction model due to its compatibility with KASCADE data. The CURVED option was chosen for primaries incident at large zenith angles and the UPWARD option for albedo particles. The energy cut for the electromagnetic component was set at 300 *MeV* since it is adequate to get all the hadrons produced by photon interactions while saving a significant amount of computing time. The code was installed with the SLANT option to study the longitudinal shower development. This data is used to determine the neutron creation and propagation while above 50 MeV.

CORSIKA ignores neutrons with energies less than 50 *MeV*. To determine the propagation and thermalization of neutrons below this energy, we use MCNP. MCNP contains high resolution neutron cross sections which are superior to other similar Monte

Carlo simulators [*Hagmann et al.* 2007]. CORSIKA outputs the longitudinal distribution of particles within the atmosphere, including neutrons above 50 *MeV*. This can be used to find the location in the atmosphere where individual neutrons pass below 50 *MeV*. For each of these locations we simulated neutrons with kinetic energy of 50 *MeV* and angle equal to the primary angle in MCNP. The atmosphere was modeled in ~100 bins corresponding to 10 *g cm$^{-2}$* column depth each. The density and size of these bins were chosen using *US Standard Atmosphere* [1976]. Flux tallies with order of magnitude neutron energy bins were set every 10 *g cm$^{-2}$* of column depth. Data obtained from the tallies of different angles at a given primary energy is then averaged by sin θ weight to give results for an isotropic distribution of primaries. (Although not all incoming distributions will be isotropic, due primarily to the geomagnetic field, we are primarily concerned with HECRs, which will be largely unaffected. The sheer volume of data prevents showing a distribution for all energies and all angles, and tallying an average over angles will be a much better approximation for most uses than giving results for a single angle of incidence.) This weighted average of neutron tallies is entered into the lookup tables for primary kinetic energies from 10 *GeV* – 1 *PeV*.

For the primary kinetic energy range 1 *MeV* – 10 *GeV*, MCNPx was used. MCNPx is the extended version of MCNP, allowing use at higher energies. MCNPx is better suited to lower energy interactions than CORSIKA, allowing for more reliable data. These simulations were performed in a way identical to the higher energy calculations, without the addition of the second step mentioned.

Identical simulations were performed using MCNPx for comparison in the energy overlap range from 1 *GeV* to 100 *GeV*. Within this range, the results of MCNPx were within statistical error of results from CORSIKA. This is consistent with other work showing comparisons between Monte Carlo simulations [*Hagmann et al.* 2007].

3. RESULTS

Over the range of energies examined, the cosmogenic neutron production increases with primary energy (Figure 2). Neutron production scales linearly with primary energy above the threshold for atmospheric neutron spallation of a few *MeV*. As the neutron production scales linearly with primary energy, the average neutron production is dependent on the total amount of energy deposited above production threshold in the atmosphere through cosmic ray interactions, and not on the high energy spectrum of the primaries.

Although the total neutron production efficiency does not significantly increase with primary energy, there is a large variation in the altitude at which the neutrons thermalize. High energy primaries produce more extensive showers freeing neutrons at lower altitudes. This creates the energy dependence of altitude distribution displayed in Figure 1.

Although these factors greatly change the magnitude of the neutron flux, the terrestrial neutron energy spectral shape remains generally invariant with respect to primary energy (Figure 3). This is to be expected due to the large number of collisions with atmospheric nuclei. Due to the invariance of the resultant spectral shape, it is probably impossible to derive the primary spectrum based solely on the neutron spectrum detected. This is consistent with the work of other researchers, which shows that neutron fluxes and GLEs scale to the magnitude of solar events, but not the spectra of individual events [*Plainaki et al.,* 2007].

4. USING THE LOOKUP TABLES

The lookup tables produced in this work are contained within auxiliary material and made freely available at: http://kusmos.phsx.ku.edu/~melott/crtables.htm .The lookup tables are organized three different ways: by primary kinetic energy, by column density, and by neutron energy. Each set contains a number of tables equal to the number of bins of that variable; 91, 94, and 10 respectively. Results display the number of neutrons per *eV* for a given bin. Primary kinetic energy is divided into 10 logarithmic bins per order of magnitude and labeled with units *log eV,* using "p" in place of a decimal point ("2p5.txt" corresponding to $10^{2.5}$ *eV*, etc.). Column density is given in units of *g cm$^{-2}$* with bins of size 10 *g cm$^{-2}$*. There are ten neutron energy bins, with labels of units *log eV*. Primary spectra from astrophysical sources often contain units of *particles m$^{-2}$ sr$^{-1}$ s$^{-1}$ GeV$^{-1}$*. The differential spectrum of neutrons can be calculated using such a spectrum by multiplying the primary spectrum by the corresponding table values and summing over the desired primary kinetic energy and neutron energy bins.

$$\frac{d\Phi_n}{dE_n} = \sum_p \frac{d\Phi_p}{dE_p} \Delta E_p \frac{dN_n}{dE_n} \qquad (1)$$

Where $\frac{d\Phi_n}{dE_n}$ refers to the differential neutron flux. Assuming the differential primary spectrum ($\frac{d\Phi_p}{dE_p}$) is in units of *particles m$^{-2}$ sr$^{-1}$ s$^{-1}$ GeV$^{-1}$*, the differential neutron flux will be given in *particles m$^{-2}$ sr$^{-1}$ s$^{-1}$ eV$^{-1}$*. Note that this primary spectrum will also be latitude dependent due to the variation of geomagnetic cutoff. $\Delta E_p$ refers to the size of primary energy bins used within our tables. Our tables provide $\frac{dN_n}{dE_n}$ for different primary energy ranges, column densities, and neutron energies in units of *particles eV$^{-1}$*. The total neutron flux can be by multiplying these individual results by their corresponding neutron energy bin size before summing.

Statistical and systematic errors were calculated for our results. Statistical errors vary depending on bin size, however in all cases the standard deviation of the mean is less than 10% of the total bin value. These errors are concentrated in regions of low neutron number, making results near the neutron production threshold the least well known. Systematic error is introduced through latitude, geomagnetic and seasonal variation. These effects can be corrected for by modulating the primary spectrum before convolution. Although systematic errors are present within CORSIKA and MCNPx, these errors are small in comparison to other sources of uncertainty. With a properly modulated cosmic ray spectrum, systematic errors are dominated by the use of proton exclusive simulations. As cosmic rays contain heavier nuclei, results will underestimate neutron production. At least 95% of this systematic error can be eliminated by use of a primary spectrum in units of *particles m$^{-2}$ sr$^{-1}$ s$^{-1}$ nucleon$^{-1}$*. For these purposes, alphas, the second-dominant species in cosmic ray shower, will count as four particles each. This will reduce the systematic error to inconsequentiality in comparison to statistical error. Although systematic errors from heavy nuclei are in principle unknown due to unknown cross sections, they prove to be small using this approximation as table results are found to match neutron monitor measurements. Our results have been tested against the neutron monitor measurements of Goldhagen et al. [2004]. This was accomplished by using the spectrum of Seo et al. [1991] as measured by LEAP below 100 *GeV*, and a differential flux proportional to E$^{-2.7}$ above 100 *GeV*. These spectra were then cut to exclude the particles with energies less than the geomagnetic cutoffs given by Goldhagen et al. [2004]. The resulting neutron flux calculations match well to these measurements, as both display the same two peaked neutron spectrum shape, and the flux of neutrons < 10 *MeV* agree to better than 20% as shown in table 2. Our measurements give fluxes less than measurements on average. This discrepancy is due to our approximation of treating helium nuclei and heavy nuclei as a number of protons equal to the number of nucleons in the nucleus. Other discrepancies arise from statistical error, numerical error due to bin size, atmospheric humidity variation, primary spectrum variations, and systematic uncertainties within the simulation code.

**Table 1.** Comparison of Results with Neutron Monitor Measurements[a]

| **Cutoff** (*GeV*) | **Atmospheric Depth** (*g cm$^{-2}$*) | **Neutron Tables Flux for E < 10 *MeV*** (*particles cm$^{-2}$ s$^{-1}$*) | **Measurement from Goldhagen et al. [2004]** (*particles cm$^{-2}$ s$^{-1}$*) | **% Error** |
|---|---|---|---|---|
| **11.6** | 53.5 | 1.08 | 0.91 | 19% |
| 0.8 | 56 | 5.92 | 7.3 | 19% |
| 0.7 | 101 | 6.47 | 7.3 | 11% |
| 4.3 | 201 | 2.28 | 2.57 | 11% |
| 2.7 | 1030 | 0.00755 | 0.0086 | 12% |

[a] Measurements taken from Goldhagen et al. [2004]

5. DISCUSSION

Cosmic ray induced neutron flux remains a threat to airplane flight crews, especially at high latitude and altitude. In the event of an increased HECR flux, effects would be prevalent at lower altitudes. These effects would be very small at ground level in most cases. Any event of energy sufficient for producing large ground level neutron radiation will be accompanied by a greatly increased and more penetrating muon flux [*Atri and Melott*, 2011a]. For this reason, biological effects of ground level cosmic ray secondary radiation will always be dominated by its muonic component. Events such as supernovae and gamma ray bursts [However, see *Abassi et al.*, 2012] could increase the HECR flux for an extended period of time, instigating or encouraging a mass extinction. Increased cosmic rays for several Myr could account for mass extinction periodicity if the HECR flux were increased periodically. A periodicity in biodiversity has been shown to exist within the fossil record [*Melott and Bambach*, 2011], with the driving mechanism still unknown [However, see *Melott et al.*, 2012]. Our results show neutron radiation doses to be small in comparison to muon radiation dose at ground level. Even a large SPE, such as the Carrington event of 1859, would only increase the neutron flux on the ground by ~ 5%. This remains very small in comparison to muon radiation dose during the same period. Although neutrons have little biological impact at ground level, they remain a threat at higher altitudes. Additionally, the soft error rate of solid state devices is affected by cosmic ray induced neutrons as well. As no simple method has existed for calculation of HECR induced neutron flux, we have developed lookup tables that can be used for primaries ranging from 1 $MeV - 1\ PeV$. Radiation dose can be calculated from these tables using appropriate biological effectiveness.

## 6. ACKNOWLEDGMENTS
We are grateful for helpful comments from two anonymous referees.We are grateful for helpful comments from two anonymous referees.

## 7. FIGURE CAPTIONS

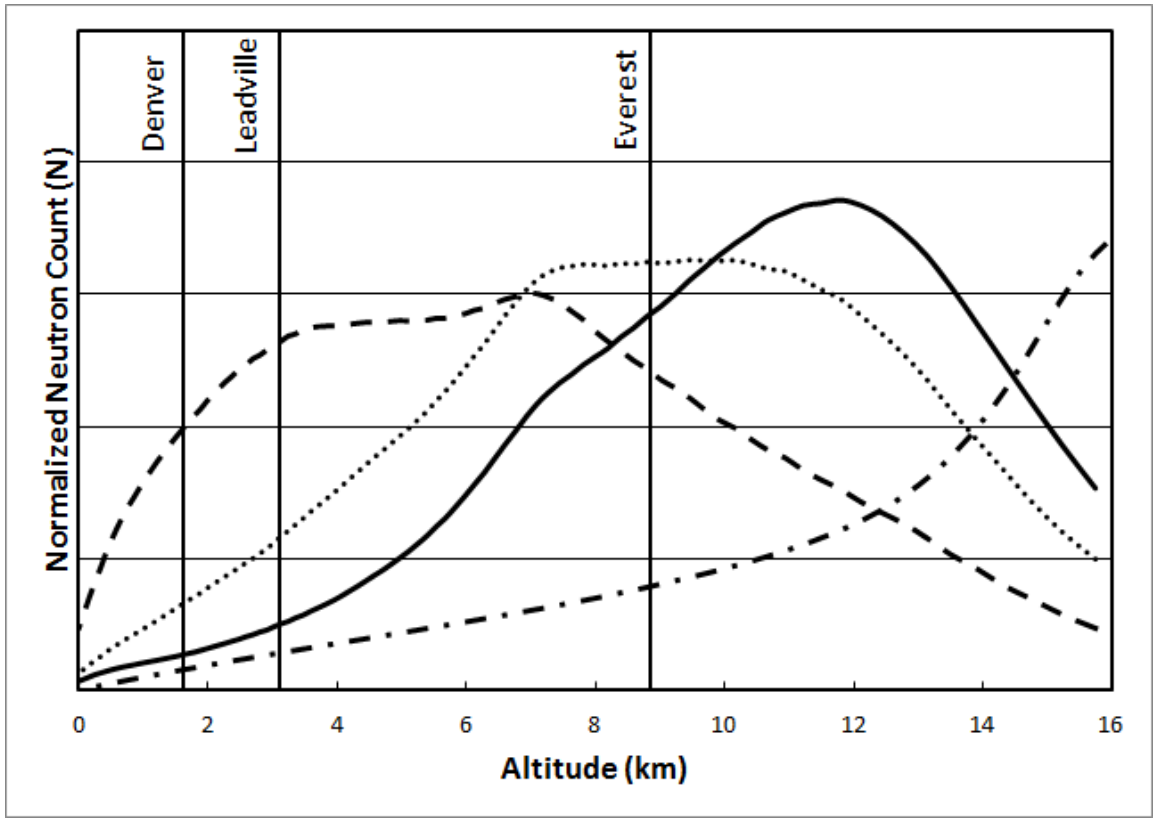

Figure 1: Normalized cosmogenic neutron count per primary as a function of altitude above sea level. Neutron count is normalized by dividing by the total number of neutrons produced in each shower. The solid line represents a 10 GeV primary, with the dotted line representing 10 TeV primary, and dashed line representing 1 PeV primary. The present neutron distribution as measured by Goldhagen et al. [2004] is given as a dash-dot line for comparison. Vertical lines correspond to commonly referenced altitudes: Denver, Colorado; Leadville, Colorado; and Mount Everest. High energy primaries produce neutrons at lower altitudes (See Figure 2) due to their secondaries having energies sufficient for further secondary production.

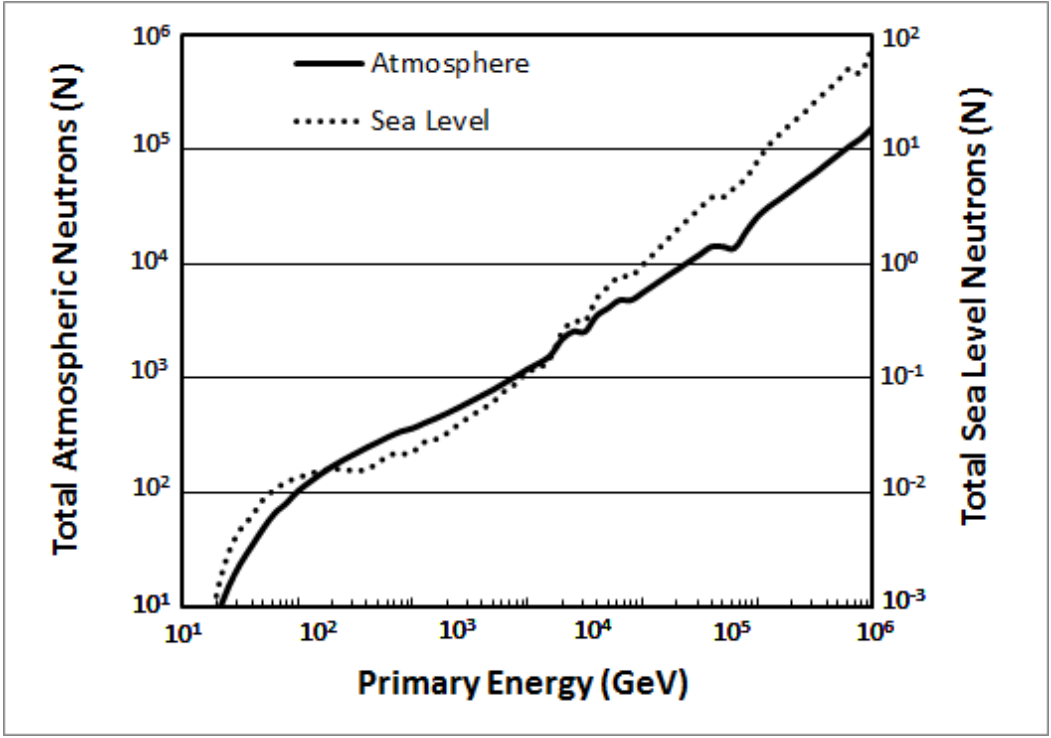

Figure 2: Total number of neutrons (of all energies) produced in the atmosphere per primary as well as neutrons which reach sea level. Atmospheric neutron number is plotted referencing the left-hand axis, with sea level neutron number referencing the right-hand axis. As shown, the growth rate of neutrons within the atmosphere is slower than at sea level. This is due to higher energy primaries depositing more energy directly on the ground without producing neutrons in the atmosphere.

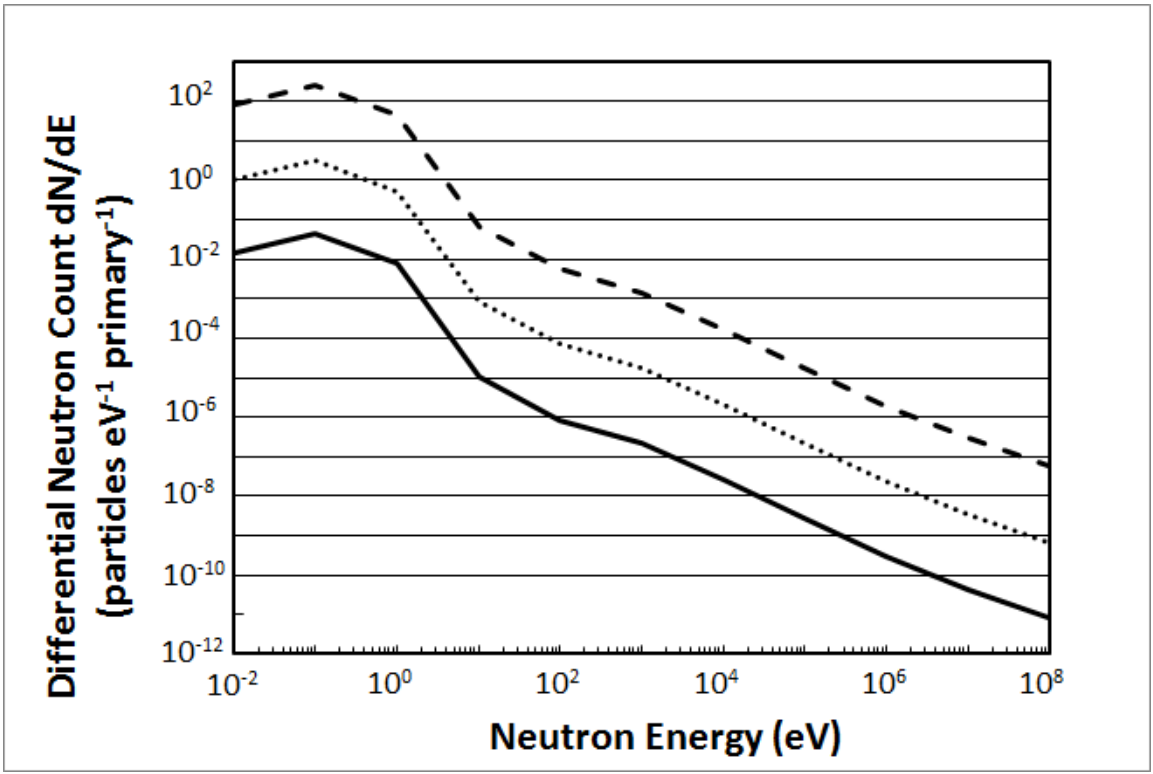

Figure 3: Differential neutron count at sea level as a function of neutron energy produced by 10 GeV (solid), 10 TeV (dotted), and 1 PeV (dashed) primaries. No neutrons reach ground level from primaries below ~1 *GeV*. Neutron energy is approximated with order of magnitude bins as with the table data. The spectral shape is similar for all energies due to the thermalization process.